\documentclass[aps,preprint,letterpaper,groupeaddress,showpacs]{revtex4}
\usepackage{amsmath,amssymb}
\usepackage{epsfig}                       \usepackage{graphicx}
\usepackage{psfrag}                       \usepackage{color}
\usepackage{psfig}                        \usepackage{amsbsy}
\usepackage{amssymb}

\begin{document}
\bibliographystyle{apsrev}
\title{First Principles Study of Work Functions of Double Wall Carbon Nanotubes}
\author{Bin Shan}
%\email{bshan@stanford.edu}
\affiliation{Department of Applied Physics, Stanford University, Stanford CA, USA, 94305-4040}
\author{Kyeongjae Cho}
\email{kjcho@stanford.edu}
\affiliation{Department of Mechanical Engineering, Stanford University, Stanford CA, USA, 94305-4040}
\date{\today}

%************************************************************************************
\begin{abstract}
Using first-principles density functional calculations, we
investigated work functions (WFs) of thin double-walled nanotubes
(DWNTs) with outer tube diameters ranging from $1nm$ to $1.5nm$. The
results indicate that work function change within this diameter
range can be up to 0.5 $eV$, even for DWNTs with same outer diameter. This
is in contrast with single-walled nanotubes (SWNTs) which show negligible
WF change for diameters larger than $1nm$. We explain the WF change and related charge redistribution in DWNTs using charge
equilibration model (CEM). The predicted work function variation of
DWNTs indicates a potential difficulty in their nanoelectronic
device applications.
\end{abstract} \pacs{31.15.Ar, 72.80.Rj, 73.30.+y} \maketitle

Double-walled nanotubes (DWNTs) have attracted considerable
attention due to their novel electronic and mechanical properties\cite{Wei}. Compared to single-walled nanotubes (SWNTs), they have higher
mechanical stiffness and greater thermal stability which may be of
benefit in field emission devices\cite{Son}. Their double shell
structure offers the possibility of shielding the inner tube from
external perturbations\cite{Endo2}. Such advantages make them very
interesting for certain device applications.

Utilization of DWNTs in early days were largely hindered by the
unwanted carbon nanomaterials generated at the same time with DWNTs
during fabrication processes. Over past few years, substantial
improvements have been achieved in fabrication and purification
techniques of DWNTs. Synthesis of high-quality DWNTs in large scale
has been realized using high temperature pulsed arc discharge
technique\cite{Toshiki} and catalytic chemical vapor
deposition\cite{Flahaut}. In a more recent experiment, production of
DWNTs with experimental yields more than $95\%$ has been
reported\cite{Endo}. Thin DWNTs has also been successfully
fabricated by by filling $C_{60}$ molecules in SWNTs followed by
electron beam irradiation, yielding DWNTs with diameter distribution
around $13\sim14\AA$\cite{Smith}. In addition, it has now become
possible to identify atomic correlations between adjacent graphene
layers using high-resolution transmission electron
microscopy\cite{Hashimoto}, enabling an unambiguous classification
of DWNTs. All these point toward a possible application of DWNTs in
nanoelectronic devices.

Recently, the potential use of DWNTs as channel materials for
field-effect transistors (FETs) has been successfully demonstrated
\cite{Takashi}, with better subthreshold swing factor as compared to
SWNTs. Also, due to different inner/outer tube combinations and
inter-layer interactions, DWNTs-based FET yields richer transistor
characteristics\cite{WangS}. For further large-scale integration of
DWNT-based FETs, an understanding of DWNT work function is
essential, since even work function change on the order of $0.1eV$
may lead to substantial changes in device
characteristics\cite{JingGuo}.

The encapsulation of molecules such as $C_{60}$ inside SWNTs are known to modify
the electronic properties of carbon nanotubes\cite{Youngmi}. Due to
their concentric double layer structures, ultra small diameter tubes can be encapsulated as the inner shell of DWNTs\cite{YLMao}. Nanotubes with
diameters $\sim 4\AA$ have already been observed in confined
channels or as the inner shell of multiwall
nanotubes\cite{ZMLi,Qin,Wang}. The inner tubes in thin DWNTs usually
feature high surface curvature and are shown to have large variations in work function\cite{BSHAN}. There have been
some previous theoretical investigations on charge redistribution in
selected DWNTs\cite{YLMao,Yoshiyuki}. However, as far as we know,
there have been no systematic study on the work functions of
DWNTs and it remains unclear how the encapsulation of ultra small inner tubes affect the work functions of DWNTs.

In this study, we report first-principles work function calculations
for two series of DWNTs consisting of either (m,m)@(n,n) pairs or
(m,0)@(n,0) pairs ($m<n$), with the first chiral index representing
the inner tube and the latter representing the outer tube. The
diameters of the DWNTs we studied range from $1nm$ to $1.5nm$,
within which work function variation is most significant. It was
found that the work function variations can be up to $0.5eV$, even
for DWNTs with similar outer diameter. This can be a significant
value in device physics. The origin of work function modulations in
DWNTs, as well as the charge redistribution are shown to be
correlated with the type of the inner shell, and can be
qualitatively understood within the framework of CEM. Some issues in
the characterization and measurements of thin DWNTs are also
discussed.
%Doping of the outer shell can be observed
%in (m,0)@(n,0) DWNTs with ($m<7$).

The work function was calculated using the standard procedure by taking the difference between the vacuum level $\phi$ and the Fermi level $E_f$. The vacuum level $\phi$ is determined from the
average potential at the center of the vacuum region where it approaches a constant. We have excluded the slowly decaying exchange-correlation part of the potential in the vacuum region to achieve better convergence\cite{Leung}.
$E_f$ was placed at the midgap in the case of a semiconducting DWNT. The calculation was done using Vienna Ab-initio Simulation
Package (VASP)\cite{VASP} within local density approximation (LDA).
30 k-points were used along the nanotube's one dimensional Brillouin
zone which was tested to give good convergence. In all simulations,
an orthorhombic unit cell with cross-sectional dimensions of
$\left(35 \times 35\right) \AA^2$ was used. With this unit cell size,
the minimum separation between periodic tube images was $>18\AA$,
which ensures the corresponding work function converges to a constant value.
Kohn-Sham single-electron wave functions were expanded by $285768$
planewaves with an energy cut-off of $286.6 eV$. For the LDA
exchange-correlation potential, we have used a functional form
fitted to the Monte Carlo results of homogeneous electron
gas\cite{Ceperley}. Conjugate-gradient method was used for both
electronic structure calculation and geometry optimization. The
DWNTs were assumed to be fully relaxed when the
force on each atom was less than $0.05\mathrm{eV}/\AA$.

\begin{figure}
\includegraphics[width=3.0in]{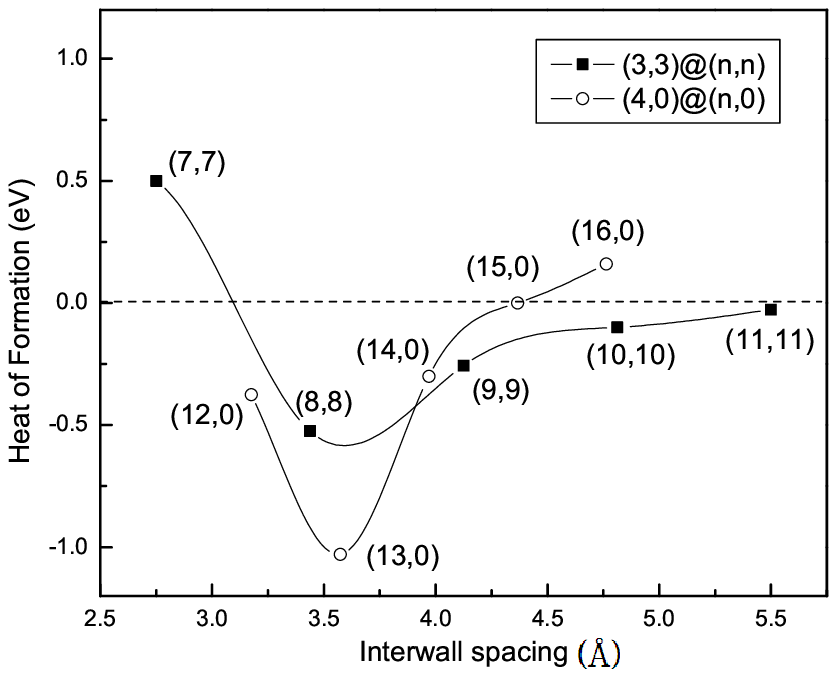}
\includegraphics[width=3.0in]{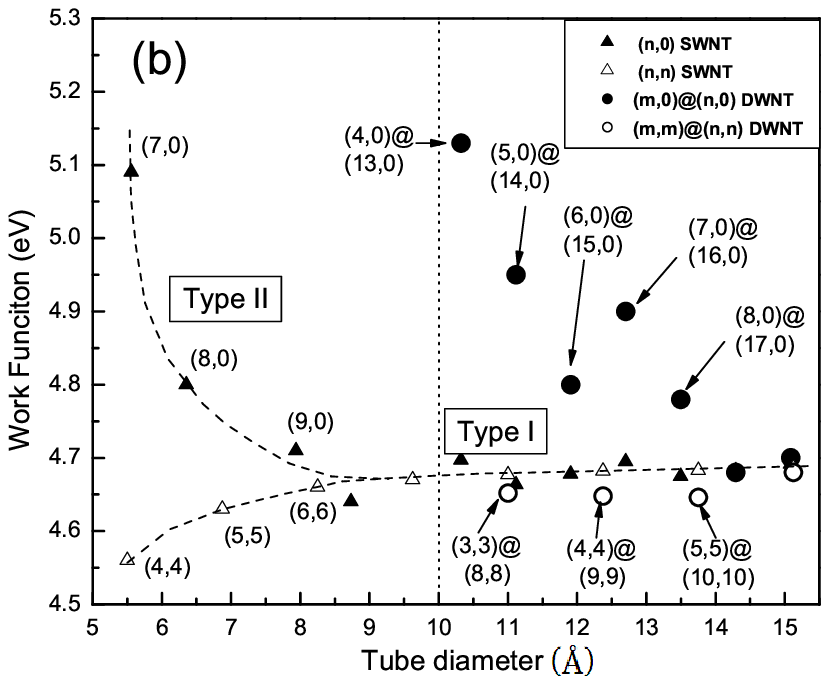}
\caption{(a)Heat of formation per unit cell for the DWNTs. solid square for (3,3)@(n,n) DWNTs and empty circle for (4,0)@(n,0) DWNTS. (b)Work functions for DWNTs(dots) and SWNTs(triangles) of different diameters}
\label{HeatofFormation}
\end{figure}

We first discuss the stability and optimal inter-wall spacing for
DWNTs. The energetics with respect to inter-wall spacing was calculated by
fixing the inner tube diameter and varying the diameter of the outer
tube. Fig. \ref{HeatofFormation}(a) shows the heat of formation for
selected DWNTs [$\Delta E=E(DWNT)-E(SWNT)_{outer}-E(SWNT)_{inner}$]
with different inter-wall spacing.  It was found that most
energetically favorable outer tube for (3,3) and (4,0) tubes are
(8,8) and (13,0) tubes, respectively. For the two series of DWNTs we
studied within diameter range, the most stable inner-outer
combination is consistently (m,m)@(m+5,m+5) for armchair pairs and (m,0)@(m+9,0) for zigzag pairs.
The optimal inter-wall spacings corresponding to such combinations
are close to the inter-layer spacing of graphite($3.35\AA$),
possibly due to structural similarities between graphene sheet and
nanotubes. This value of inter-wall spacing is consistent with other
first-principles calculations\cite{Charlier} and experimental values
from scanning tunneling microscope measurements\cite{Sattler}. Even
though in real fabrication processes, DWNTs with non-optimal
inter-wall spacing are also expected to exist, those with optimal
inter-wall spacing is more energetically favorable and would prevail
in the final product. Moreover, similar correlation to the inner
tube chirality was seen in DWNTs with non-optimal inter-wall
spacing. Thus, in the following, we focus our discussion on work
functions of DWNTs with optimal inter-wall spacing.

Figure \ref{HeatofFormation}(b) summarizes work functions of various
DWNTs with optimal inter-wall spacing, plotted against the diameter
of the outer tube. Also shown on the graph are work functions of
SWNTs, indicated by the dotted line. As can be seen from the graph,
there is essentially no diameter dependence of work functions for
SWNTs larger than $1nm$ (Type I SWNT). However, different work
functions of DWNTs can be observed in the diameter range from $1nm$
to $1.5nm$, for up to $0.5eV$. More specifically, (m,0)@(n,0) DWNTs
show a general decrease trend in work functions while those of
(m,m)@(n,n) pairs show little variation. This indicates that the
work function difference is primarily due to the presence of the
inner ultra small diameter tube (Type II SWNT), which are known to
have substantial differences in work functions\cite{BSHAN}.
In particular, when the inner shell of the DWNTs are of (m,0) zigzag
type, the overall work function of the DWNT was increased due to the
higher work function of the inner tube.

The qualitative feature of the work function variations in DWNTs can
be understood by the CEM, which has been successfully applied to
molecular dynamics simulations\cite{Goddard}. Generally speaking,
when two materials systems with different chemical potentials (minus
of work function) comes into contact, chemical potentials are equalized
by electron transfer from higher chemical potential system to
lower potential system. The total electrostatic energy of such a
system can be written, neglecting higher order terms, as

\begin{eqnarray}
E_{tot}(q_1,q_2,...,q_N)&=& \sum_{i}^N \left(E_i^0+(\frac{\partial
E}{\partial q_i}) q_i+\frac{1}{2}(\frac{\partial^2 E}{\partial
q_i^2})q_i^2\right)+\frac{1}{2}\sum_{i\neq
j}V_{ij}(q_i,q_j) \\
&=&\sum_{i}^N \left(E_i^0+\chi_i^0
q_i+\frac{1}{2}J_i^0q_i^2\right)+\frac{1}{2}\sum_{i\neq
j}V_{ij}(q_i,q_j)
\end{eqnarray}

Where $E_i^0$ is the energy of neutral atom $i$, $-eq_i$ is the
excess electrons on atom $i$, $\chi_i^0$ is the electronegativity (minus of work function),
$J_i^0$ is the atomic hardness, and $V_{ij}$ is the Coulombic interaction
between atoms $i$ and $j$. In the context of DWNTs, we can view the DWNT as a giant
'molecule', with inner and outer tube being two artificial 'atoms'
$A$ and $B$. Taking the derivative of $E_{tot}$ with respect to
$q_i$, we arrive at the following equations for chemical potential and charge transfer:
\begin{subequations}
\begin{equation}
\chi_A(q)=\chi_A^0+J_A^0q+\frac{q}{2C}
\end{equation}
\begin{equation}
\chi_B(-q)=\chi_B^0-J_B^0q-\frac{q}{2C}
\end{equation}

\end{subequations}
where $C=\frac{2\pi\varepsilon}{ln(b/a)}$ is the unit length capacitance
between two coaxial cylinders, with $a$ and $b$ being the inner and outer tube diameter, respectively. The dielectric constant of the
nanotube is set to 1\cite{Leonard}. Under equilibrium, the
electrochemical potential of the inner tube and outer tube must
equal. By equating $\chi_A(q)$ and $\chi_B(-q)$, we have the following
first-order solution for the final chemical potential and charge
transfer for DWNTs:

\begin{equation}
q=\frac{\chi_B^0-\chi_A^0}{J_A^0+J_B^0+1/C}
\end{equation}
\begin{equation}
\chi=\frac{J_B^0+\frac{1}{2C}}{J_A^0+J_B^0+\frac{1}{C}}\chi_A^0+\frac{J_A^0+\frac{1}{2C}}{J_A^0+J_B^0+\frac{1}{C}}\chi_B^0
\end{equation}

It can be seen that the direction and amount of charge transfer is
directly related to the work function difference between the inner
and outer tubes, which is confirmed by subsequent charge
redistribution analysis. The final work function of DWNT is a linear
combination of work functions of the inner and outer tubes. Due to
the small inter-layer spacing and the resulting large $1/C$ term,
the work function of DWNT can be roughly approximated by the average
of work function of inner and outer tubes. Table \ref{DWNTWF} lists
work functions predicted by the CEM as well as those from
first-principles calculations. The general trend of work function
change predicted from CEM is in good agreement with the
first-principles calculations, with the largest error $\sim 0.2eV$.

\begin{figure}
\includegraphics[width=3.0in]{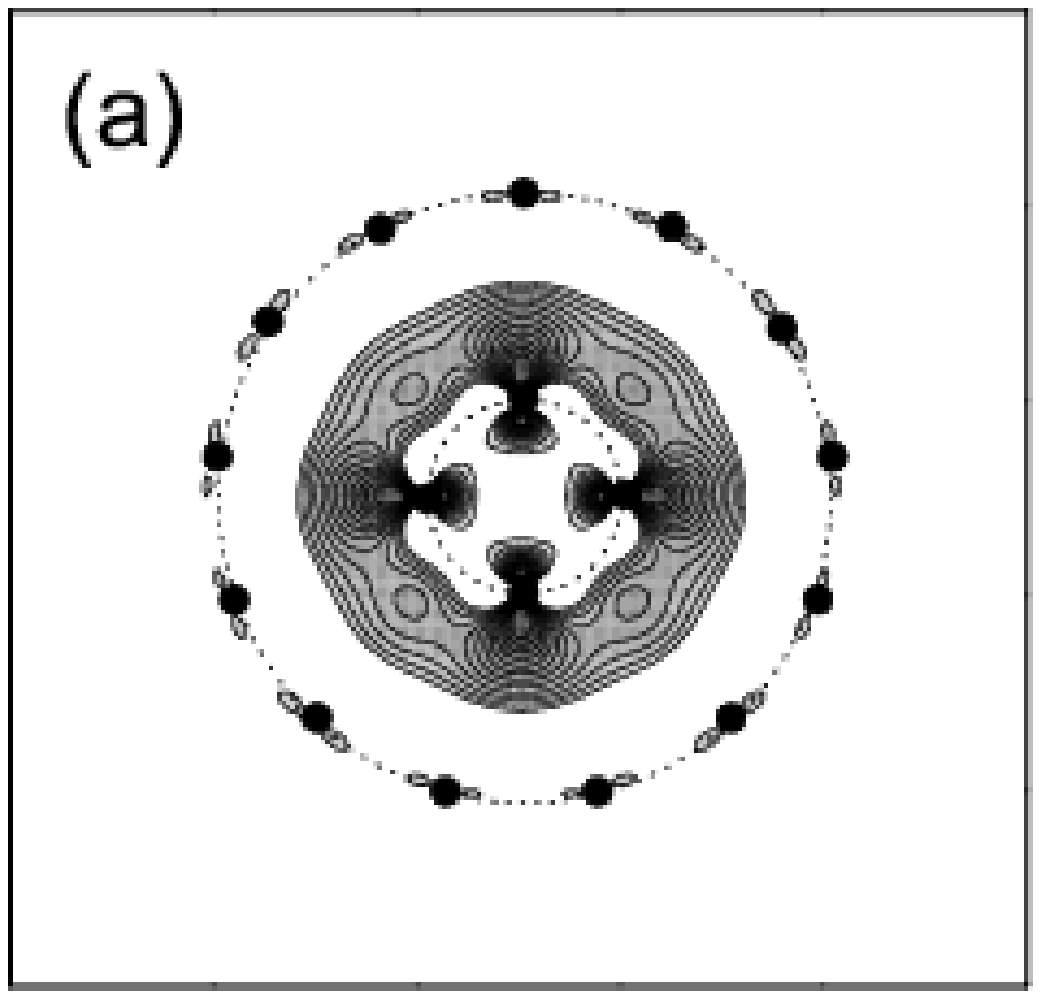}
\includegraphics[width=3.0in]{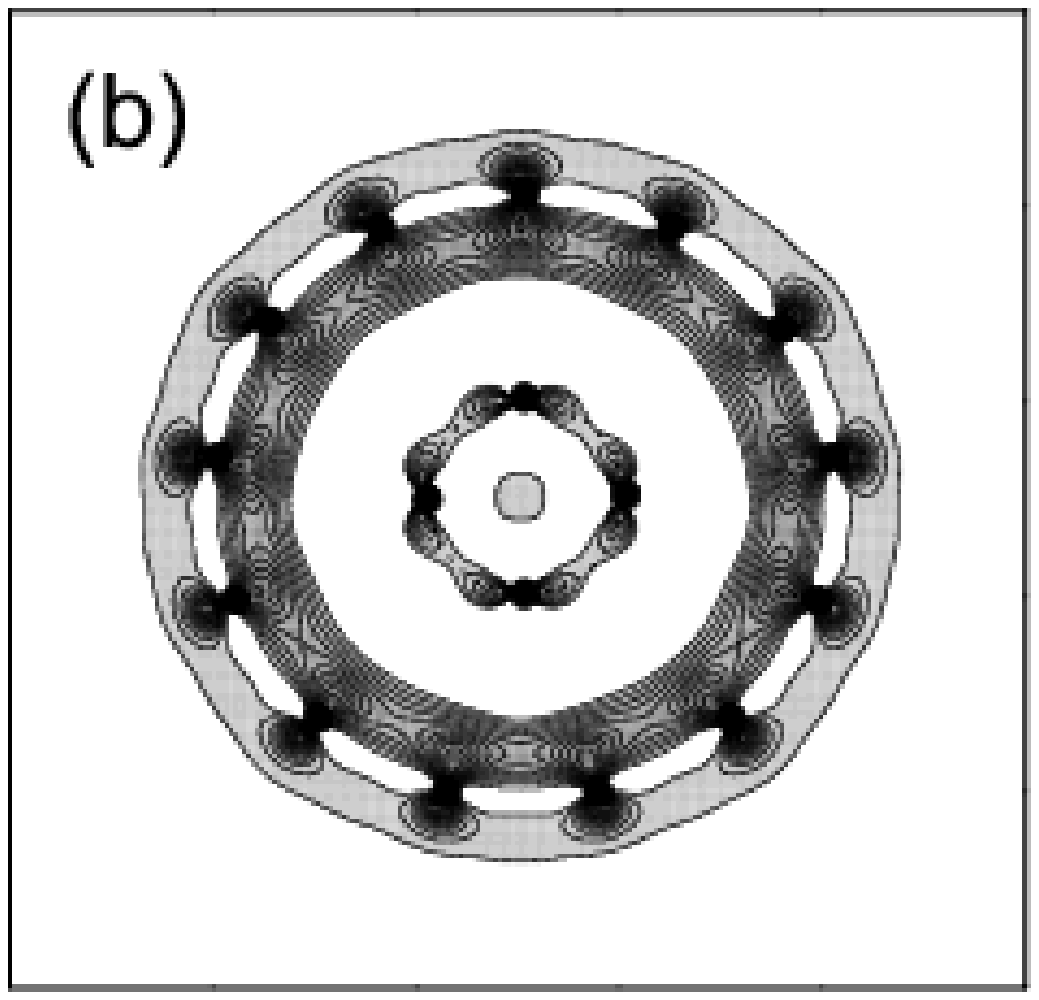}
\includegraphics[width=3.0in]{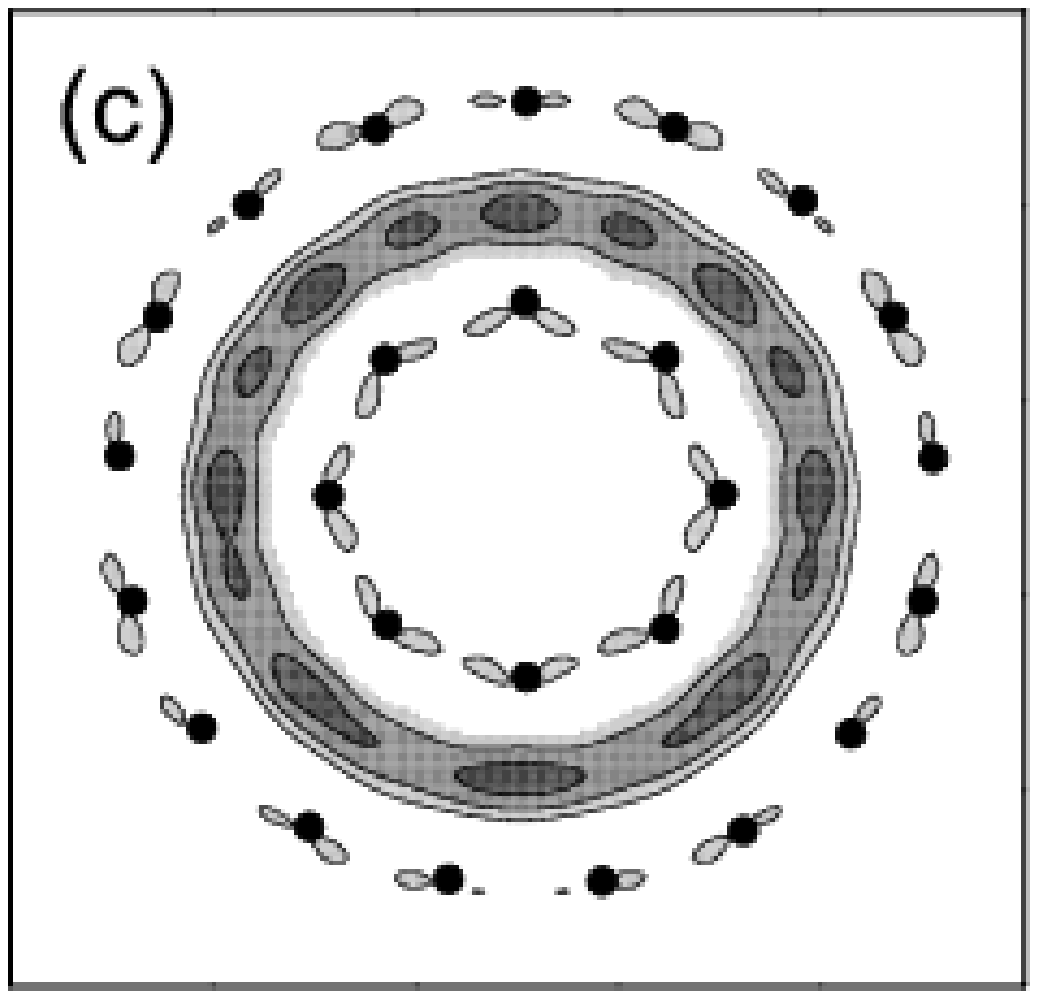}
\includegraphics[width=3.0in]{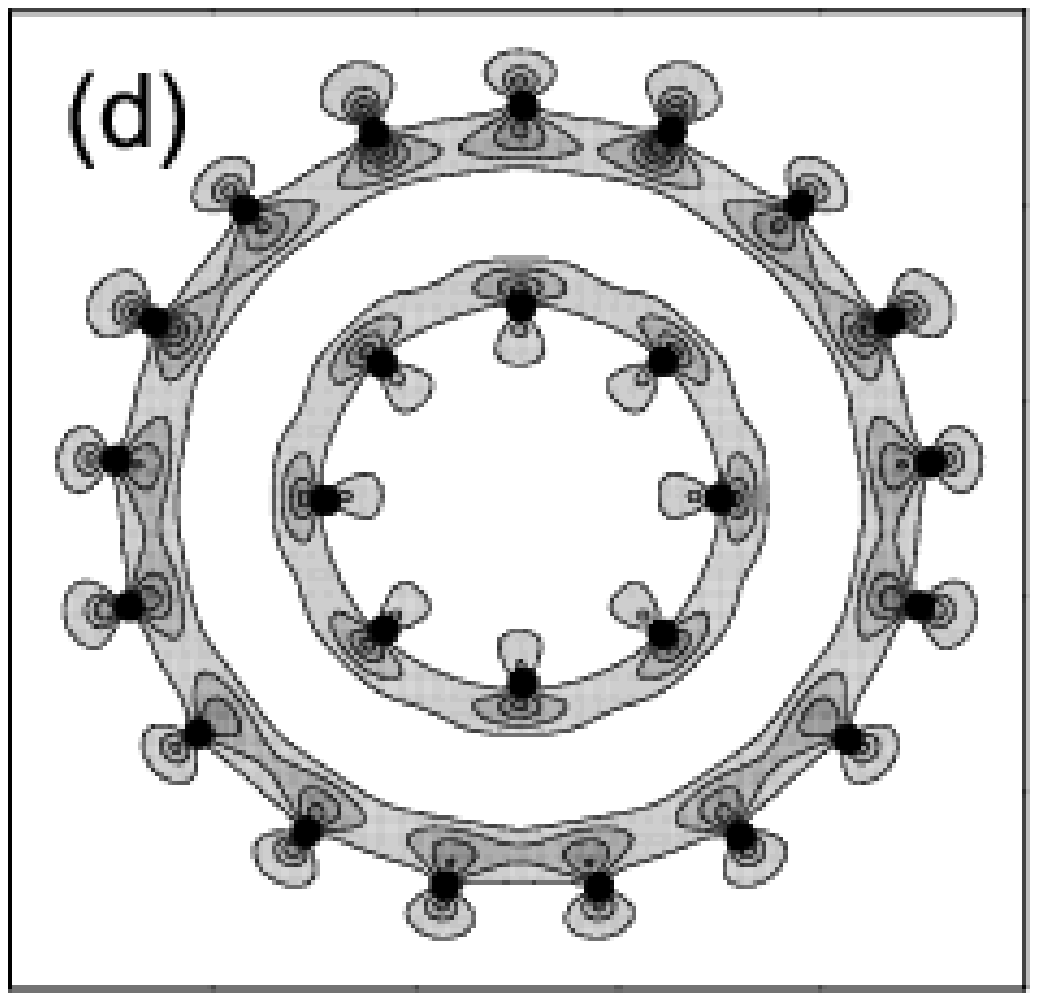}
\includegraphics[width=3.0in]{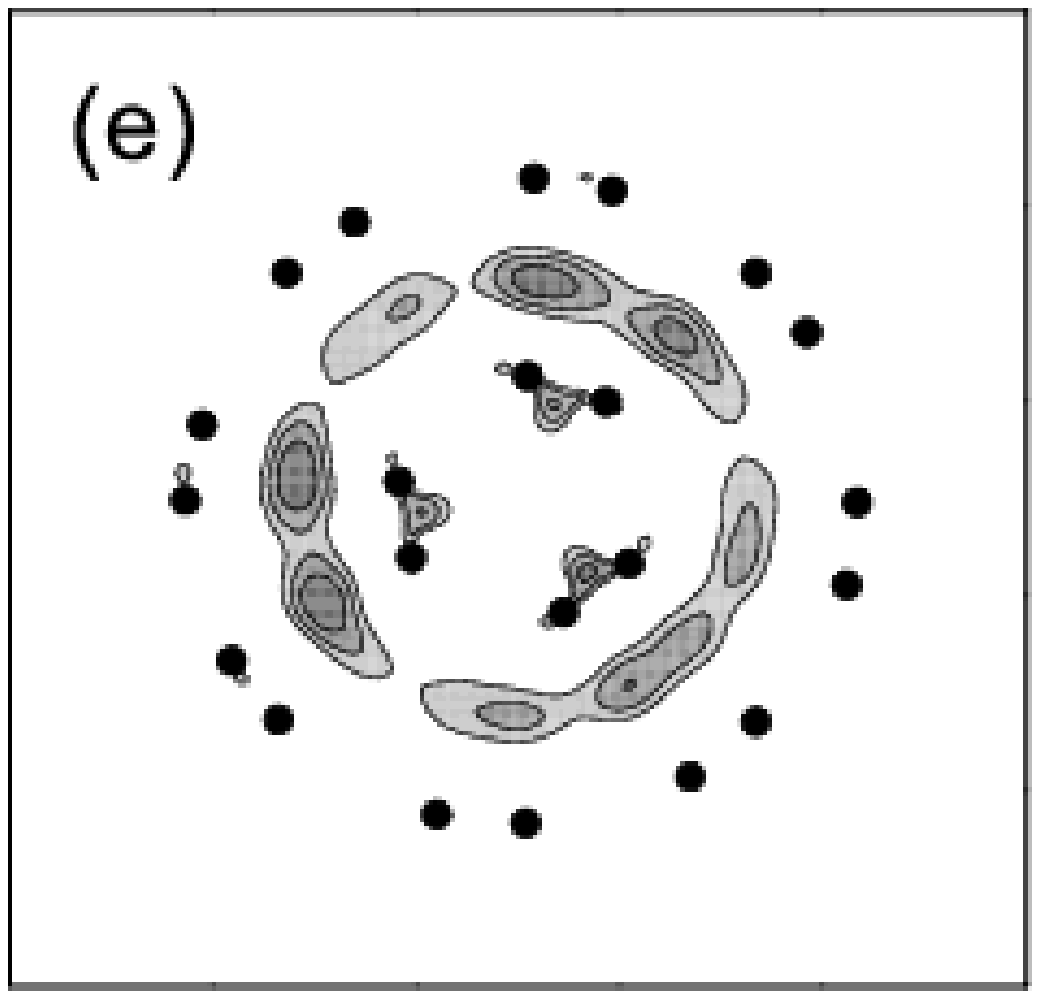}
\includegraphics[width=3.0in]{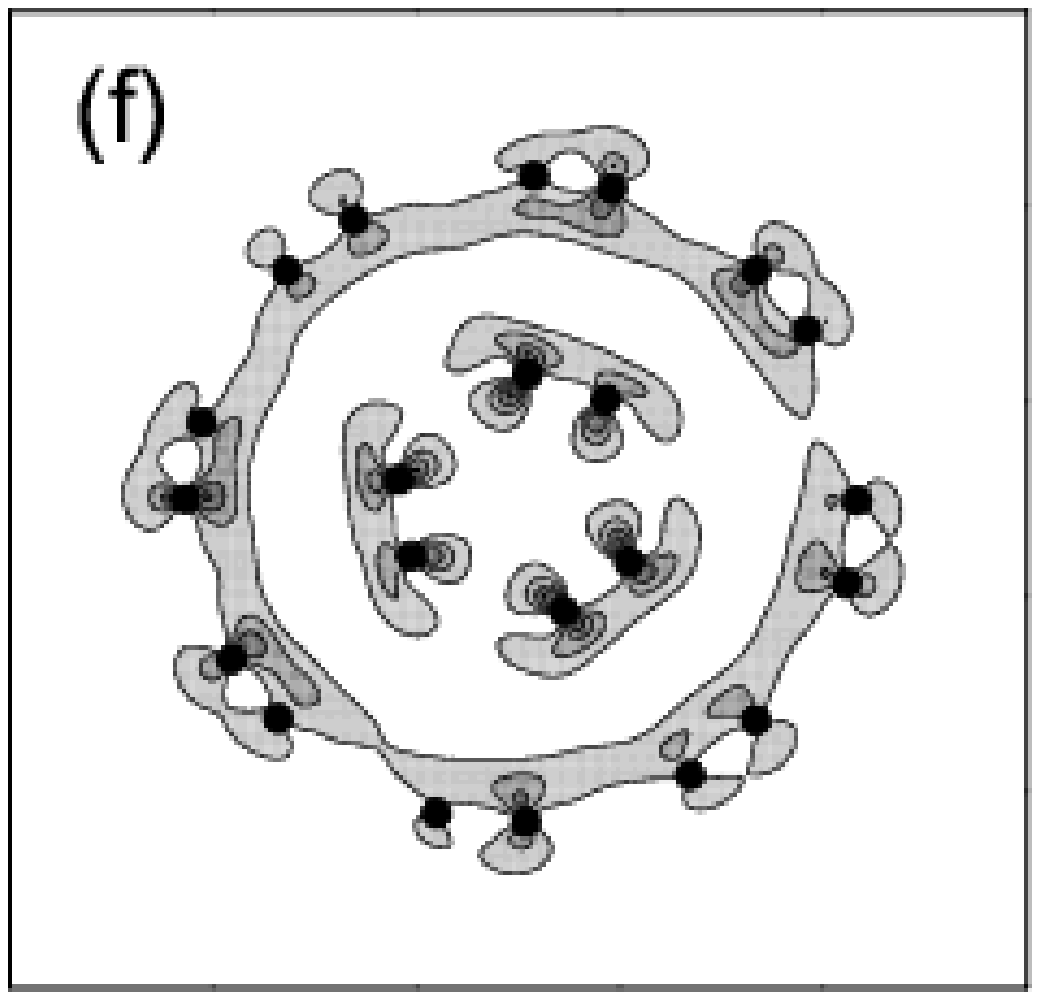}
\caption{Contour plots of charge accumulation (left column) and depletion region (right column) for (4,0)@(13,0), (8,0)@(17,0) and (3,3)@(8,8) DWNTs, respectivly. The contour plots are shown in a plane normal to the tube axis, with black dots indicating positions of carbon atoms. The contour value spacings are set to be equal.}
\label{ChargeTransfer}
\end{figure}

The charge redistribution in DWNTs are calculated by subtracting the
charge distribution of isolated inner and outer tubes from the
self-consistent charge distribution, $\Delta
\rho=\rho(DWNT)-\rho(inner)-\rho(outer)$.
Fig.\ref{ChargeTransfer}(a,c,e) and (b,d,f) shows the charge
accumulation regions and charge depletion regions, respectively.
 In (4,0)@(13,0)
DWNT, the work function of the inner tube is $\sim 1eV$ higher than
that of the outer tube. Due to this large difference in work
function, the amount of charge transfer is also more significant
than in other DWNTs, as indicated by the dense contour lines in Fig.
\ref{ChargeTransfer}(a) and (b). The direction of charge transfer
can also be clearly identified. The accumulated charge mainly
locates on the inner (4,0) tube region rather in in the empty
inter-tube region. Such charge transfer pattern has also been
observed in (5,0)@(14,0) tubes. This indicates that provided large
enough work function difference, the direction of charge transfer
can be predicted from SWNT work functions, consistent with one's
physical intuition. In the case of (8,0)@(17,0) and (3,3)@(9,9)
DWNTs, the work function difference of inner and outer tube is not
as large and the direction of charge transfer is more subtle due to
the presence of inter-wall interaction. The depleted charge from the
$\pi$ electron system accumulates in the intertube region, with
characteristic similar to the interlayer state in graphite systems.
Such interlayer state in DWNTs has also been observed in other
first-principles studies\cite{Yoshiyuki}. The amount of charge
transfer is also smaller as evident by the sparse contour lines on
Fig. \ref{ChargeTransfer}(c-f).

\begin{figure}
\includegraphics[width=3.0in]{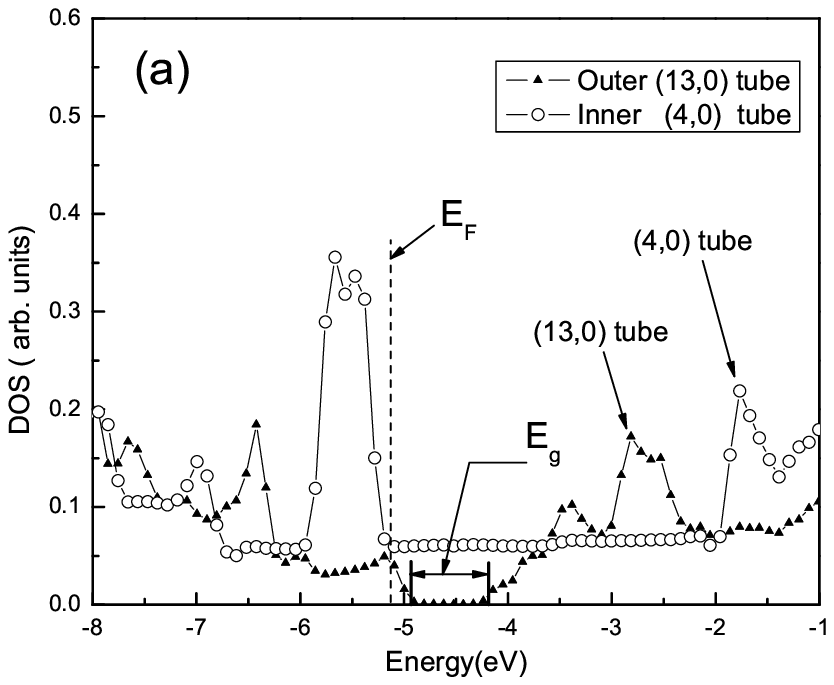}
\includegraphics[width=3.0in]{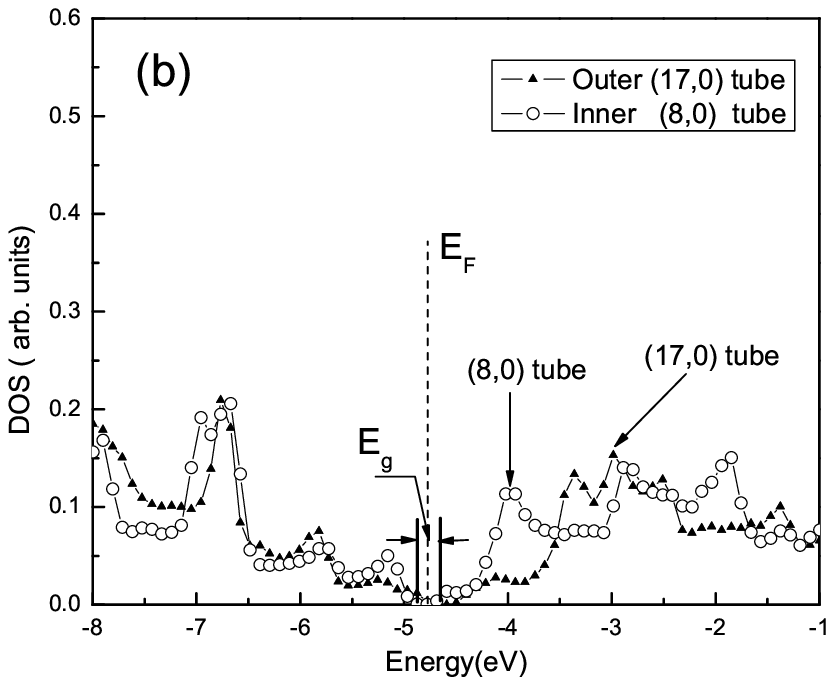}
\includegraphics[width=3.0in]{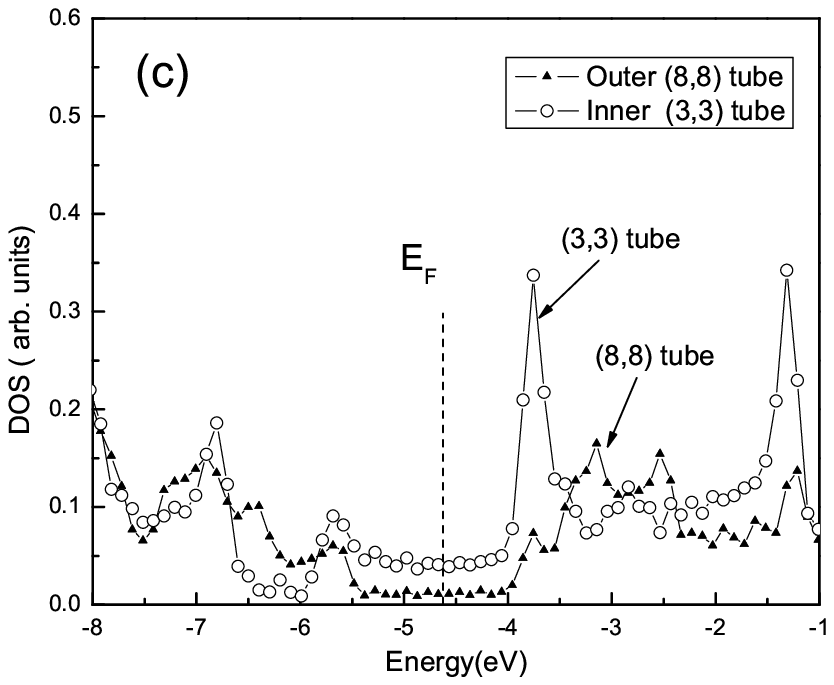}
\caption{Site-projected DOS for (4,0)@(13,0), (8,0)@(17,0) and
(3,3)@(8,8) DWNTs respectively. Empty circles for the outer tube and
filled triangles for the inner tube.} \label{LDOS}
\end{figure}

Finally, we discuss some possible issues in the characterization of
thin DWNTs. Fig. \ref{LDOS} shows the site-projected density of states (DOS), i.e. local DOS integrated within spheres of Wigner-Seitz radius centered on each ion, for the
above three DWNTs. In (m,0)@(n,0) DWNTs such as (4,0)@(13,0), the outer tube is semiconducting with low WF and  the inner tube is metallic with high WF due to $\sigma-\pi$ hybridization\cite{BLASE}. When they form DWNT, substantial
charge transfer from the outer tube to the inner tube leads to hole doping of
the outer tube. Even though the band gap of the semiconducting outer tube is largely retained despite of the inter-wall interaction, the Fermi level is below its valence band edge, giving rise to a finite DOS at the
fermi energy (Fig. \ref{LDOS}(a)). For those DWNTs with ultra small zigzag SWNTs as inner
shell, their work functions are thus expected to be $0.2\sim0.5eV$ higher
than larger diameter DWNTs. As the DWNT diameter gets larger, such
as for a (8,0)@(17,0) pair, the work functions of the inner and
outer tube are comparable and their individual electronic structures
are less perturbed. The DWNT is a semiconductor with a reduced band gap as compared to band gaps of both the outer and inner tube (Fig. \ref{LDOS}(b)). (3,3)@(8,8) DWNT presents an interesting case.
Even though it has almost the same diameter as (4,0)@(13,0) one
and is also metallic, its electronic structure around the fermi
level is considerably different. Due to the presence of inter-wall
interaction, states at the fermi level are mostly localized in the
inner tube with tails extending to the outer shell (Fig. \ref{LDOS}(c)). Its work
function is about $0.5eV$ lower than that of (4,0)@(13,0) DWNT, due to the higher WF of the inner tube\cite{BSHAN}.
These differences are likely to affect the transport and scanning
tunneling microscope measurements of DWNTs.

In conclusion, we have systematically studied the work functions of
two series of DWNTs. It was found that the work function differences
of DWNTs in a narrow diameter distribution ($1.0nm\sim1.5nm$) can be
up to $0.5eV$. The general trend of the work function change and
charge redistribution are successfully explained by the CEM. These
findings provides useful insight into the work function change
in DWNTs which may be used to develop new ways of engineering electronic structures
of DWNTs.
\begin{acknowledgments}
    This work is supported by NSF grant on Network for Computational Nanotechnology (NCN). Part of the calculations are done at San Diego Supercomputing Center.
\end{acknowledgments}

\begin{table}
\caption{Work Functions of selected DWNTs}
\begin{tabular}{l|c|c|c|c|c}
\hline\hline
DWNT   &      Tube diameter   & Outer Shell  &  Inner Shell WF    &   WF from CEM  &   WF from DFT \\
        &   ($\AA$)         &   WF (eV)*    &   (eV)*       &              (eV)         &   (eV)   \\
\hline
(4,0)@(13,0)       &     10.32    &  4.70   &    5.95  &  5.32  &  5.13 \\
\hline
(5,0)@(14,0)     &     11.11   &   4.66   &  5.10 &     4.88    & 5.00  \\
\hline
(6,0)@(15,0)     &     11.91   &   4.68   &  4.99 &      4.84   & 4.80  \\
\hline
(7,0)@(16,0)     &     12.70   &   4.69   &  5.10 &      4.90   & 4.90  \\
\hline
(8,0)@(17,0)      &    13.49  &  4.69 &  4.80 &    4.74     &4.77     \\
\hline
(3,3)@(8,8)     &      11.00   &  4.68  &  4.50  &    4.59     &4.64    \\
\hline
(5,5)@(10,10)   &      13.75       &  4.68  &    4.63   &      4.66     & 4.64     \\
\hline\hline
\end{tabular}\\
*work functions of SWNTs from Ref\cite{BSHAN}
\label{DWNTWF}
\end{table}

\end{document}